\documentclass{osa-article}
\journal{oe}

\articletype{Research Article}

\begin{document}

\title{Theoretical analysis of the role of complex transition dipole phase in XUV transient-absorption probing of charge migration}

\author{Yuki Kobayashi,\authormark{1,*} Daniel M. Neumark,\authormark{1,2} and Stephen R. Leone\authormark{1,2,3}}

\address{
\authormark{1}Department of Chemistry, University of California, Berkeley, CA 94720, USA\\
\authormark{2}Chemical Sciences Division, Lawrence Berkeley National Laboratory, Berkeley, CA 94720, USA\\
\authormark{3}Department of Physics, University of California, Berkeley, CA 94720, USA}

\email{\authormark{*}ykoba@stanford.edu} 

\begin{abstract}
We theoretically investigate the role of complex dipole phase in the attosecond probing of charge migration.
The iodobromoacetylene ion (ICCBr$^+$) is considered as an example, in which one can probe charge migration by accessing both the iodine and bromine ends of the molecule with different spectral windows of an extreme-ultraviolet (XUV) pulse. 
The analytical expression for transient absorption shows that the site-specific information of charge migration is encoded in the complex phase of cross dipole products for XUV transitions between the I-$4d$ and Br-$3d$ spectral windows.
Ab-initio quantum chemistry calculations on ICCBr$^+$ reveal that there is a constant $\pi$ phase difference between the I-$4d$ and Br-$3d$ transient-absorption spectral windows, irrespective of the fine-structure energy splittings.
Transient absorption spectra are simulated with a multistate model including the complex dipole phase, and the results correctly reconstruct the charge-migration dynamics via the quantum beats in the two element spectral windows, exhibiting out-of-phase oscillations.
\end{abstract}

\section{Introduction}
Recent progress in laser technology has enabled real-time tracking of photochemical dynamics from the most fundamental point of view, i.e., electron motion \cite{Krausz09}.
One such phenomenon is known as charge migration, in which a coherent superposition of electronic states drives oscillatory motion of electrons from one end of a molecule to another \cite{Woerner17,Tuthill20}. 
The characteristic few-electronvolt energy spacing of molecular valence orbitals dictates that coherent electronic motion can occur even before nuclear motions set in (e.g., 1 eV of energy spacing corresponds to an oscillation period of 4.1 fs).
Observation of charge migration has been a central topic in attosecond science \cite{Leone14,Young18}, and a number of experimental \cite{Calegari14,Kraus15,Mansson21,Barillot21} and theoretical \cite{Kuleff05,Dutoi11,Jenkins16,Bruner17,Hollstein17,Despre18,Mauger18,Cho18,Jia19,Chen20,Folorunso21,Golubev21,Khalili21,Yong21} studies have reported results the last few years.
There are both exciting perspectives and unanswered questions regarding charge migration.
Potential implications of electronic coherence in photochemisry remain a topic of debate, such as efficient charge transfer in light-harvesting antenna \cite{Engel07} and selective bond dissociation in photoionized peptide molecules \cite{Weinkauf96}.
The mechanisms of electronic decoherence due to nuclear motions or many-body interactions have been investigated theoretically and will benefit from more experimental results \cite{Woerner17}.
Laser-control of charge migration is predicted to be attainable \cite{Golubev15}, and realization of this can be a fundamental step toward attochemistry \cite{Remacle98,Nisoli19,Merritt21}.

In order to see charge-migration dynamics in real time, one needs a spectroscopic probe that has, in addition to a sub-femtosecond temporal resolution, the capability to resolve site-specific electron density.
One way to achieve this is to utilize the localized nature of inner-shell electrons, which can be accessed by high-energy photons in the x-ray or extreme-ultraviolet (XUV) regime.
Among the various spectroscopic methods in this energy regime \cite{Chergui17,Kraus18}, the technique considered here is transient absorption spectroscopy, wherein an optical pulse triggers photochemical reactions and an XUV pulse imprints the dynamics in the core-to-valence absorption spectra \cite{Geneaux19}.
Combined with attosecond XUV pulses produced by high-harmonic generation, the method has been successfully applied to measure electronic coherences in atoms \cite{Goulielmakis10,Wirth11,Kobayashi18} and molecules \cite{Kobayashi20DBr,Kobayashi20Br2}.
The site specificity of XUV absorption can be most efficiently utilized by accessing multiple tagging elements in a molecule with a broadband XUV light source \cite{Kobayashi19IBr}, with the potential to achieve a panoramic reconstruction of charge migration.

Simulation of charge migration and XUV absorption therein requires quantum-mechanical treatments, as charge migration is inherently a quantum-mechanical process.
One of the possible ways is to solve the time-dependent Schr{\"o}dinger equation for a finite number of electronic states, and compute the absorption signals from a coherent superposition system as a Fourier transformation of dipole oscillations \cite{Wu16}.
What distinguishes this method, compared to an absorption spectrum that is constructed as a simple sum of the absolute squares of the transition dipole moment (or oscillator strength), is that it can account for quantum interference between multiple transition pathways.
Indeed, characteristic quantum beats that are induced by a quantum interference from a pair of valence states that end up at a common final state is the key to attosecond transient-absorption probing of electronic coherence \cite{Goulielmakis10,Santra11}.
A question arises when considering the aforementioned scheme of multielement probing of charge migration: what determines the relative timing of the quantum beats when multiple final states are present, especially when these final states belong to different elemental spectral windows?

Here, we highlight the role of the complex transition dipole phase in understanding XUV absorption probing of charge migration.
The complex dipole phase is lost when absorption signals or transition probabilities are simply considered to be proportional to oscillator strengths.
However, as recently shown for solid-state high-harmonic generation \cite{Jiang17,Jiang18} and attosecond probing of light-dressed states \cite{Yuan19}, or as traditionally well known in two-pathway excitation experiments (e.g., $\omega$-3$\omega$) \cite{Zhu95,Seideman98}, the complex dipole phase is not negligible when multiple transition pathways interfere.
The target molecule of the present study is iodobromoacetylene (ICCBr).
The I-$4d$ and Br-$3d$ core-level absorption edges lie in the XUV regime ($\sim{}50$ eV and $\sim{}65$ eV, respectively), which can be accessed by a typical high-harmonic generation setup \cite{Timmers16,Li20}.
A molecule of similar structure, iodoacetylene (ICCH), was already used for charge migration studies via high-harmonic spectroscopy \cite{Kraus15,Jiang18}, and the additional bromine atom in ICCBr enables the possibility of multi-element probing of charge migration in attosecond transient absorption.
We first examine analytical expressions for the attosecond transient absorption signals from a coherent system, and show that the complex phase of cross transition dipole products determines the beat phase for each probe state.
Ab-initio calculations are performed for the electronic structure of ICCBr$^+$, and the complex dipole phases are revealed to be constant within each of the I-$4d$ and Br-$3d$ windows irrespective of the spin-orbit coupling, ligand-field effects, and probe direction.
Attosecond XUV absorption spectra are simulated by solving the time-dependent Schr\"{o}dinger equation within a multistate model, and the manifestation of the complex dipole phase is analyzed.

\section{Computational methods}
\subsection{Electronic structure of ICCBr$^+$}
The electronic structure of ICCBr$^+$ is computed by using the spin-orbit general multiconfigura-tional quasidegenerate perturbation theory (SO-GMC-QDPT) implemented in a developer version of GAMESS-US \cite{Schmidt93,Nakano02,Zeng17,Kobayashi19HBr}.
Relativistic model-core potentials and basis sets of triple-zeta quality (MCP-TZP) are used in the calculations \cite{Miyoshi98,Sekiya01}.
A fixed linear structure of the molecule with bond lengths of C-I 1.99 \AA{}, C-C 1.18 \AA{}, C-Br 1.79 \AA{} is used \cite{Almenningen76}. 
The active space contains 11 valence orbitals and the 10 core-level orbitals (i.e., I-$4d$ and Br-$3d$), and a total of 42 single-hole configurations are included in the state-average calculations.
The transition dipole moments including their complex phase are obtained for all combinations of the valence and core-excited states.
Note that the wavefunctions of the valence and core-excited states are obtained in a single set of calculations, and the transition dipole moments are calculated straightforwardly by using the Slater-Condon rule.
In order to accurately compute the spin-orbit couplings of the halogen atoms, effective nuclear charges of 71.37 and 41.40 are used for the iodine and bromine atoms, respectively.
In addition, the spin-orbit matrix coupling constants for the Br-$3d$ and I-$4d$ shells are scaled by factors of 0.577 and 0.665, respectively \cite{Kobayashi19IBr}.

\subsection{Multi-state simulation of core-to-valence absorption spectra}
Core-to-valence absorption spectra of ICCBr$^+$ are simulated by solving the time-dependent Schr\"{o}dinger equation for a multi-state model,
\begin{align}
i\frac{\partial}{\partial t}\Psi(t) = \left[ H_0 - \frac{i\Gamma}{2} - \mathbf{d} \cdot \mathbf{E}(t) \right] \Psi(t).
\end{align}
In Eq. (1), $\Psi(t)$ is a column vector that contains the complex coefficients for the electronic states, $H_0$ is a field-free Hamiltonian whose diagonal elements correspond to the state energies, $\Gamma$ is a diagonal matrix for the autoionization lifetime of the core-excited states, $\mathbf{d}$ is a transition-dipole matrix, and $\mathbf{E}(t)$ is the laser electric field.
The model consists of two valence states, X $^2\Pi_{3/2}$ and A $^2\Pi_{3/2}$, and the ten I-$4d$ and Br-$3d$ core-excited states.
The energy and dipole moments are obtained from the SO-GMC-QDPT calculations, the autoionization lifetimes of all the core-excited states are assumed to be $\Gamma=100$ meV \cite{Nahon92PRA,Sullivan96}, which corresponds to a $1/e$ lifetime of 6.6 fs.
The XUV absorption spectra are obtained by calculating the single-atom absorption cross section,
\begin{align}
\sigma(\omega) \propto \omega \mathrm{Im}\left[ \frac{d(\omega)}{E(\omega)} \right],
\end{align}
where $d(\omega)$ and $E(\omega)$ are the dipole moments and applied laser field, respectively, that are Fourier transformed from the time domain to the frequency domain \cite{Wu16}.

\section{Results}
\subsection{State-resolved core-to-valence absorption spectra}
Figure 1 shows the calculated XUV absorption of ICCBr$^+$ from 4 different electronic states, X $^2\Pi_{3/2}$, A $^2\Pi_{3/2}$, B $^2\Pi_{3/2}$, and C $^2\Sigma_{1/2}$ \cite{Heilbronner70}.
The absorption spectra are constructed as a sum of the oscillator strengths convoluted with a 0.5-eV Gaussian broadening.
The neutral molecule has an electronic configuration of $1\sigma^2 1\pi^4 2\pi^4 3\pi^4$, and the inset images show the molecular orbitals that yield the main single-hole configuration of each electronic state.
The results show that the different electronic states make unique fingerprints in the XUV absorption spectra.
Furthermore, the site specificity of core-to-valence absorption is demonstrated; the 3$\pi$ and 2$\pi$ orbitals have comparable weights from the I-$5p$ and Br-$4p$ orbitals, and the absorption signals appear both in the I-$4d$ and Br-$3d$ windows for the X $^2\Pi_{3/2}$ and A $^2\Pi_{3/2}$ states [Figs. 1(b) and 1(c)].
The 1$\pi$ orbital, on the other hand, has little contribution from the I-$5p$ orbital, and the absorption signals from the B $^2\Pi_{3/2}$ state are highly suppressed in the I-$4d$ window but are strong in the Br-$3d$ window.
Similarly, the 1$\sigma$ orbital mostly consists of the I-$5p$ orbital, and the absorption signals from the C $^2\Sigma_{1/2}$ state are more enhanced in the I-$4d$ window.
These site and state specificities of core-level absorption are what make this method an appropriate tool to probe charge migration. 

\begin{figure}[tb]
\centering\includegraphics[scale=1.0]{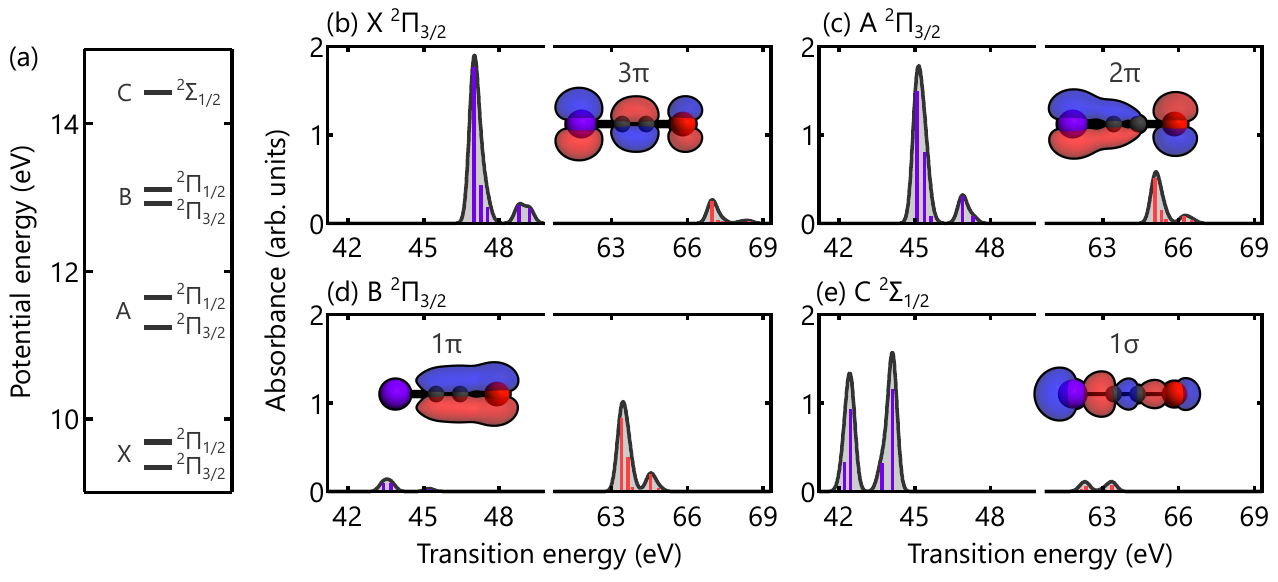}
\caption{
(a) An energy diagram of valence electronic states of ICCBr$^+$.
(b-e) Calculated core-to-valence absorption spectra of the (b) X $^2\Pi_{3/2}$, (c) A $^2\Pi_{3/2}$, (d) B $^2\Pi_{3/2}$, and (e) C $^2\Sigma_{1/2}$ states.
The inset images show the main single-hole configuration of the respective electronic states.
The site and state specificities of the core-to-valence absorption signals are shown.
}
\end{figure}

\subsection{Complex-dipole phase in analytical expression of transient absorption}
We next consider charge migration dynamics that arise from a coherent superposition of the X $^2\Pi_{3/2}$ and A $^2\Pi_{3/2}$ states of the ion.
The energy difference between the two states is 1.90 eV \cite{Heilbronner70}, and the corresponding beat period is $T=2.2$ fs.
Figure 2(a) shows the hole density motion of the charge migration, which is calculated as a coherent superposition of the $3\pi$ and $2\pi$ hole states (Figs. 1(b) and 1(c)).
At one timing, defined as $t_0$, the hole density is localized on the I atom; a half period later, the hole density migrates to the carbon and Br sites.

\begin{figure}[tb]
\centering\includegraphics[scale=1.0]{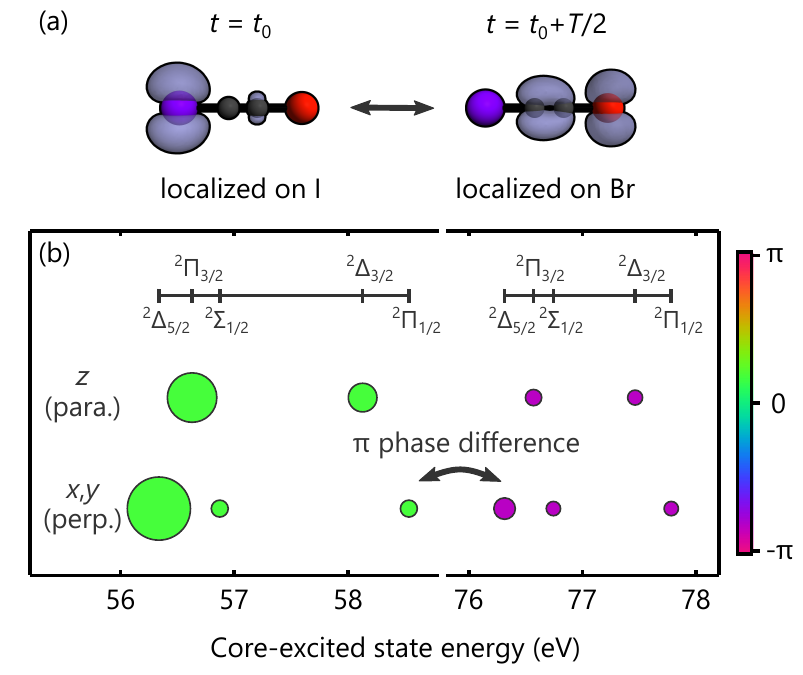}
\caption{
(a) Snapshots of the charge migration that arises from a coherent superposition of the X $^2\Pi_{3/2}$ and A $^2\Pi_{3/2}$ states.
(b) Cross products of the transition dipole moments, $d_{nF} d_{mF}^*$.
The size and color of the circles represent product magnitude and phase, respectively. 
The valence states are $n$: X $^2\Pi_{3/2}$ and $m$: A $^2\Pi_{3/2}$.
The probe states $F$ are the I-$4d$ and Br-$3d$ core-excited states.
The calculated complex phases are 0.566 radians (green) and -2.575 radians (purple) in the iodine and bromine windows, respectively, thus yielding a phase difference of $\pi$.
}
\end{figure}

In order to investigate how the core-to-valence absorption from the I-$4d$ and Br-$3d$ windows probe such hole-density motion, we start with a general analytical formula for the attosecond transient absorption spectra of a coherent atomic system, whose derivation is given in Ref. \cite{Santra11}.
The single atom response at a probe timing $t$ is expressed as,
\begin{align}
\sigma(\omega,t)=
\frac{4\pi\omega}{c} \mathrm{Im}\left[ \sum_{n,m}\rho_{nm}(t) \sum_{F} \frac{d_{nF} d_{mF}^*}{ \Delta \omega_{F,m} - \omega -i\Gamma_F/2 } \right],
\end{align}
where $n,m$ are the labels for the valence states, $F$ is the label for the core-excited states, $\rho$ is the density matrix of the targeted valence states, and $\Delta\omega$ is the transition energy.
The electronic coherence of the system, i.e., the off-diagonal density matrix element $\rho_{nm}$ ($n \neq m$), gives rise to a quantum beat with its amplitude mainly determined by a cross product of the transition dipole moments $d_{nF} d_{mF}^*$.
Specifically at the peak center $\omega=\Delta \omega_{F,m}$, the equation simplifies to
\begin{align}
\sigma(\omega=\Delta \omega_{F,m},t)=
-\frac{8\pi\Delta\omega_{F,m}}{c \Gamma_F} \mathrm{Re}\left[ \sum_{n,m}\rho_{nm}(t) \sum_{F} d_{nF} d_{mF}^* \right].
\end{align}
This equation shows that the choice of the probe state $F$, be it I-$4d$ or Br-$3d$ core-excited states, is manifested in the following three factors: (i) the XUV absorption peak width via $\Gamma_F$, (ii) the XUV absorption peak strength at line center via $\Delta \omega_{F,m}$, and (iii) the overall XUV absorption peak strength and time dependence via the cross dipole product, $d_{nF} d_{mF}^*$.
Factors (i) and (ii) only affect the static information and are not important here, whereas factor (iii) is directly related to the amplitude and phase of time-dependent spectral absorption and modulation depth that arise from charge migration.

The cross transition dipole products for the specific case of ICCBr$^+$, where $n=$ X $^2\Pi_{3/2}$ and $m=$ A $^2\Pi_{3/2}$, are calculated and shown in Fig. 2(b).
The results are shown for the parallel ($\Delta \Omega = 0$) and perpendicular ($\Delta \Omega = \pm 1$) excitations, and different sizes and colors indicate the amplitude and phase, respectively.
Note that there is an arbitrary phase offset in the cross transition dipole products and only the relative difference for different $F$ states has physical meaning.
The energy structure of the probe states $F$, i.e., the I-$4d$ and Br-$3d$ core-excited states, can be explained as follows \cite{Kobayashi19HBr}.
The single-hole configuration of $nd^{-1}$ gives rise to a $^2D$ electronic state in the atomic limit.
For the case of a linear molecule, the presence of the neighboring atoms, i.e. the molecular ligand fields \cite{Bancroft86,Cutler92,Johnson97}, causes energy splittings for the different orbital angular momenta on the order of tens of meV, giving rise to the $^2\Delta$, $^2\Pi$, and $^2\Sigma$ states.
Spin-orbit coupling is critical for $d$ orbitals, causing few-hundred meV additional energy splittings, and the electronic fine structure shows five final states, $^2\Delta_{5/2,3/2}$, $^2\Pi_{3/2,1/2}$, and $^2\Sigma_{1/2}$. 
The results in Fig. 2(b) show that within each elemental window the phase of the cross dipole products is constant despite the energy splittings caused by the ligand-field effects and spin-orbit coupling.
Furthermore, the phases between the I-$4d$ and Br-$3d$ windows exhibit a $\pi$ phase difference, both for the parallel and perpendicular probes.
This result is directly related to an intuitive picture that the multielement probe of the targeted charge migration will show out-of-phase oscillations in the I-$4d$ and Br-$3d$ element windows.

It bears mentioning that the phase difference is not always $\pi$; for example, a case of $n=$ X $^2\Pi_{3/2}$ and $m=$ B $^2\Pi_{3/2}$ (see Fig. 1) yields a phase difference of zero between the I-$4d$ and Br-$3d$ windows.
This can be understood, for lower electronic states of ICCBr$^+$, by counting the number of nodes in the molecular wavefunctions.
The number of nodes along the direction of the chemical bonds is 2, 1, and 0 for the 3$\pi$, 2$\pi$, and 1$\pi$ orbitals, respectively (see Fig. 1).
The wavefunctions at the bromine and iodine sites are in phase for the 3$\pi$ and 1$\pi$ orbitals, whereas they are out of phase for the 2$\pi$ orbital.
Due to the localized nature of the core orbitals, this local phase information of the valence orbitals is directly encoded in the core-to-valence transition dipole moments.
Returning to the case mentioned above, the transition dipole moments from the I-$4d$ and Br-$3d$ orbitals are both in phase for the X $^2\Pi_{3/2}$ and B $^2\Pi_{3/2}$ states, and thus the phase difference in the cross transition dipole products is zero.


\subsection{Multistate model transient absorption spectra}
By using the complex transition dipoles calculated for ICCBr$^+$, attosecond transient absorption spectra of the charge migration in the X $^2\Pi_{3/2}$ and A $^2\Pi_{3/2}$ states are simulated by solving the time-dependent Schr\"{o}dinger equation in a multistate model [Fig. 3(a)].
The parallel probe is considered in the following discussion, but the same conclusions are obtained for the perpendicular probe. 
Time zero is defined as when the I-$4d$ absorption is maximized, and the signals in the Br-$3d$ window (>63 eV) are multiplied by a factor of 4 to improve the visibility.

\begin{figure}[tb]
\centering\includegraphics[scale=1.0]{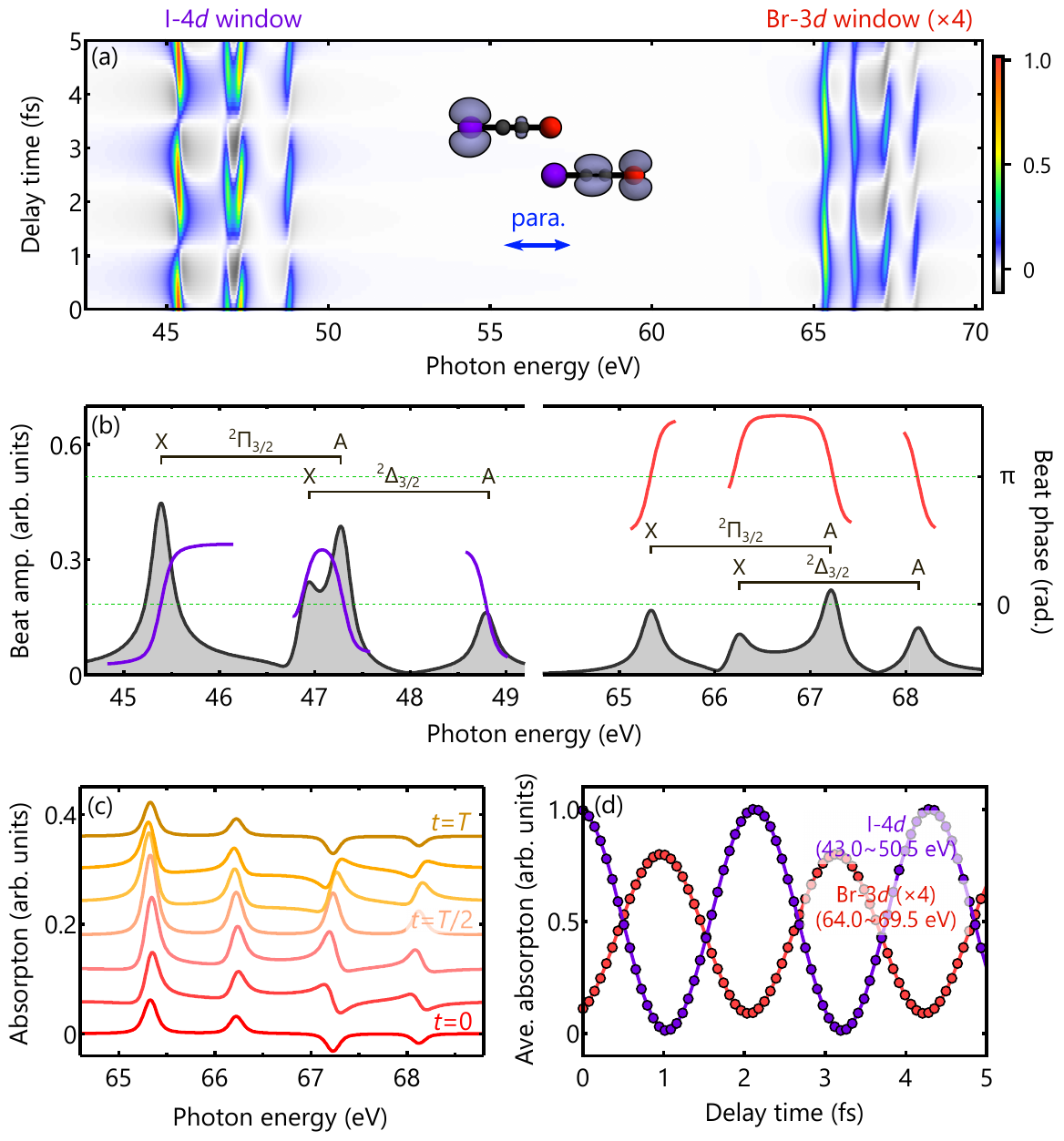}
\caption{
(a) Attosecond transient absorption spectra simulated within a multistate model including the complex dipole phase.
The probe direction is parallel to the molecular axis.
The initial state is a coherent superposition of the X $^2\Pi_{3/2}$ and A $^2\Pi_{3/2}$ states with equal weights.
The signals in the Br-$3d$ window are multiplied by a factor of 4.
(b) The amplitude (gray area, left axis) and phase (purple and red curves, right axis) of the quantum beats determined by cosine-function fitting.
It can be seen that the beat phase shows an abrupt change around the signal maxima.
(c) Snapshots of the absorption signals from $t = 0$ to $t = T$.
In addition to the signal amplitude, the line shape also exhibits periodic modulations.
(d) Average absorption in the I-$4d$ (43.0-50.5 eV) and Br-$3d$ (64.0-69.5 eV) windows.
The average signals show a phase difference of 3.0 radians, a value close to $\pi$.
}
\end{figure}

In each elemental window, there are four absorption signals that correspond to the transitions from the X $^2\Pi_{3/2}$ and A $^2\Pi_{3/2}$ states to the $^2\Pi_{3/2}$ and $^2\Delta_{3/2}$ core-excited states.
All the signals exhibit quantum beats with a 2.2-fs period, as expected from the 1.90 eV energy difference.
The beat phase can be analyzed by performing cosine-function fitting to absorption lineouts at each photon energy with a fixed period of 2.2 fs.
Figure 3(b) shows the fitted amplitude (gray area, left axis) and phase (purple and red curves, right axis) of the quantum beats in the I-$4d$ and Br-$3d$ windows.
The results show that the beat phase is not constant with XUV photon energy around the signal maxima; for example, the signal around 45.4 eV (X $^2\Pi_{3/2}$ to $^2\Pi_{3/2}$) exhibits a phase reversal versus XUV photon energy from $-\pi/2$ to $\pi/2$, and at the signal center the phase is almost exactly zero.

The beat phase is not constant because, in addition to the signal amplitude, the line shape also exhibits periodic modulations.
Figure 3(c) shows the absorption signals in the Br-$3d$ window for one beat cycle.
It can be seen that the shape of the absorption signals is changing between Lorentz and Fano lineshapes \cite{Ott13,Kobayashi17}, which is a direct outcome of the quantum interference in the core-to-valence transitions.
The time and energy dependence of the line shape as well as the spectral overlap of the absorption signals make it challenging to determine the beat phase for each individual signal.
Experimentally, it would be desirable to have a simple approach to determine the beat phase for each signal, instead of performing two-dimensional global fitting to the entire spectra with Eq. (3).
One such approach is to only examine signals that have little to no overlap with neighboring ones.
For example, the lowest signals in each element window, i.e., the X $^2\Pi_{3/2}$ to $^2\Pi_{3/2}$ transitions at 45.4 eV and 65.3 eV, exhibit beat phases of 0.01 and 3.16 radians, respectively, which are almost equal to 0 and $\pi$.
Another approach is to take a spectral average and determine one representative phase in each element window.
Figure 3(c) shows the time evolution of the averaged absorption for the I-$4d$ (43.0-50.5 eV) and Br-$3d$ (64.0-69.5 eV) windows.
This simple analysis yields a phase difference of $3.0$ radians, which is close to the expected value of $\pi$.
In either approach, it needs to be verified that the complex dipole phases are expected to be constant within the targeted energy window by performing electronic-structure calculations.

\section{Conclusions}
We examined the role of the complex transition dipole phase in the attosecond transient absorption probe of charge migration.
An example system of ICCBr$^+$ is considered, which allows for multielemental probing of electron dynamics at the molecular termini.
The analytical expression for the transient absorption signals shows that the complex phase of the cross transition dipole products plays a critical role in determining the time-dependent spectral absorption and modulation timing.
The ab-initio electronic-structure calculations show that the complex phases of the cross transition dipole products are constant within each of the I-$4d$ and Br-$3d$ windows for the X-A charge migration despite the energy splittings caused by the ligand-field effect and spin-orbit coupling.
More importantly, there is a $\pi$ phase difference between the two element windows, which corresponds to the intuitive picture of electron density migrating between the iodine and bromine ends of the molecule.
The simulated absorption spectra show that the charge migration causes periodic modulations both in the amplitude and line shape of the XUV absorption.
Spectral overlap of the absorption signals can add complexity to experimental characterization of the intrinsic beat phases, and it is suggested that using spectrally isolated signals or taking spectral averages within a targeted window can be a practical way to circumvent the problem.
Our results are limited to a multistate model with frozen nuclei, and accurate simulations of the transient absorption probe of charge migration require several additional effects such as molecular vibrations and electron-hole correlation \cite{Bruner17,Golubev21}.
Nevertheless, it is demonstrated that a multistate model can reproduce the panoramic probe of charge migration if the complex dipole phase is included in the calculation of absorption spectra, and this will make a handy tool for future experimental studies.

\begin{backmatter}
\bmsection{Funding}
National Science Foundation (NSF) (Grant No. CHE-1951317) (Y.K. and S.R.L.); U.S. Department of Energy, Office of Science, Office of Basic Energy Sciences, Gas Phase Chemical Physics Program (Grant No. DE-AC02-05-CH11231) (D.M.N.).


\bmsection{Disclosures}
The authors declare no conflicts of interest.

\bmsection{Data Availability Statement}
Data underlying the results presented in this paper are not publicly available at this time but may be obtained from the authors upon request.


\end{backmatter}

\bibliography{biblist}

\end{document}